\documentclass[12pt,oneside,a4paper]{article}
\usepackage{amssymb}
\usepackage{amsthm}
\usepackage{amsmath}
\usepackage{enumitem}
\usepackage{float}
\usepackage{wrapfig}
\usepackage{mathtools}
\usepackage{graphicx}
\usepackage{caption}
\usepackage{subcaption}
\usepackage{cases}

\usepackage{xcolor}
\usepackage[utf8]{inputenc}
\usepackage[english]{babel}
\usepackage{wrapfig}

\begin{document}

\title{A Nonlinear Formulation of Radiation Stress and Applications to Cnoidal Shoaling}
\author{Martin O. Paulsen $\cdot$ Henrik Kalisch}
\date{}
\maketitle

g

\begin{abstract}
\noindent
In this article we provide formulations of energy flux and radiation stress consistent
with the scaling regime of the Korteweg-de Vries (KdV) equation.
These quantities can be used to describe the shoaling of cnoidal waves 
approaching a gently sloping beach. 
The transformation of these waves along the slope can be described
using the shoaling equations, a set of three nonlinear equations
in three unknowns: the wave height $H$, the set-down $\bar{\eta}$
and the elliptic parameter $m$. 
We define a numerical algorithm for the efficient solution
of the shoaling equations, and we verify our shoaling formulation
by comparing with experimental data from two sets of experiments
as well as shoaling curves obtained in previous works.
\end{abstract}

\paragraph{Keywords:} Surface waves $\cdot$ KdV equation $\cdot$ Conservation laws $\cdot$ Radiation stress $\cdot$ Set-down

\section{Introduction} 

In the present work, the development of surface waves across a gently sloping bottom is in view.
Our main goal is the prediction of the wave height and the set-down of a periodic wave as it enters an area of
shallower depth. This problem is known as the shoaling problem and has a long history,
with contributions from a large number of authors going back to Green and Boussinesq.

In order to reduce the problem to the essential factors, we make a number
of simplifying assumptions. 
First, the bottom slope is small enough to allow
the waves to continuously adjust to the changing depth
without altering their basic shape or breaking up. Second, reflections are assumed to be negligible 
so that the energy flux of a wave is conserved as it transforms on the slope. Third, the period $T$
of a wave is assumed to be constant as it progresses. This last point is sometimes termed the 
conservation of waves as the number of waves in a group will remain constant during
the shoaling process.

Rather than following a wave in space and time as it propagates over the slope,
we estimate wave properties as functions of the
depth using conservation of period $T$ and energy flux, 
and a prescribed change in momentum due to the effect of the radiation stress. 
These conservation equations are due to the assumptions stated above, and they form
the basis for the description of the shoaling process as first envisioned by
Rayleigh 
\cite{Rayleigh}, and subsequently used by a number of
authors. The crucial factor which guarantees that the shoaling assumptions
are valid is that the relative change in water depth over a wavelength
is smaller than the wave steepness. This point is explained in more
detail in \cite{OsPe1, svendsen2006introduction}.

In the context of linear wave theory, the process outlined above is the
classical approach to wave shoaling and can be found in
textbooks on coastal engineering such as \cite{dean1984water,Sorensen}.
This approach depends on the assumption that the waves can be described
by a sinusoidal function of the form $\frac{H}{2} \cos(kx-\omega t)$,
where $H$ is the wave height, $k$ is the wavenumber, $\omega$
is the circular frequency, $t$ is the time and $x$ is a spatial coordinate
along which the bottom is sloping.
Recall that the wavelength is $\lambda = \frac{2 \pi}{k}$
and the wave period is $T = \frac{2 \pi}{\omega}$, and these are related
by the dispersion relation $\omega^2 = gk \tanh{(kh)}$ at local depth $h$.
In the linear shoaling theory, the conservation of energy flux is central.
The energy flux is generally formulated as the average over one period
of the energy density $E$ times the group speed $C_g$.
The energy density is given in terms of the fluid density $\rho$, 
the gravitational acceleration $g$ and the wave height $H$ as $E=\frac{1}{8} \rho g H^2$,
and the group speed is given by $C_g = \frac{d \omega}{d k}$.

If the bathymetry is sloping only gently upwards,
then it may be assumed that the waves adjust adiabatically to the changing
conditions, and that reflections and distortions are negligible.
The wave-height transformation of a shoaling wave is obtained by imposing
conservation of the energy flux $E C_g$ across the shoaling region.
If a wave measurement $H_0$, $T_0$ at some point offshore is given, then the wave height
at a point closer to shore may be computed from the conservation of energy flux
given in the form
\begin{equation}
    H  C_g = H_0 C_{g0},
    \label{linear shoaling equation}
\end{equation}
\noindent
where the group speed $C_g$ at local depth $h$ can be found from the conservation of 
the wave period $T$.
Similar considerations using the linear definition of the radiation stress
yield a formula for the set-down \cite{longuet1964radiation}.

The authors of \cite{svendsen1972}
used cnoidal functions
in connection with the linear formulation of the energy flux defined above to obtain a 
hybrid theory of shoaling.
Unfortunately, this approach led to a discontinuity in wave height at the matching point between 
the linear shoaling equation \eqref{linear shoaling equation} 
and the nonlinear theory based on the cnoidal functions. 
The problem was remedied to some degree in \cite{svendsen1977wave}
by requiring continuity in wave height effected through the use of conversion tables.
This approach led to good agreement with the experimental data, but was cumbersome
to implement, and as already noted in \cite{svendsen1977wave},
the continuity of the wave height at the matching point in the shoaling curve 
led to a discontinuity in energy flux. 

The problem can be resolved
by defining a shoaling equation based on the full water-wave problem
using the streamfunction method \cite{dean1984water} for example.
Such an approach has been documented in \cite{Dixon}.
While accurate, the streamfunction approach depends on discretizing the steady Euler equations
leading to large systems of equations which need to be solved numerically.
If the approach is to be used in connection with ocean-wave statistics,
the approach based on the steady Euler equations might be too computationally demanding.
Moreover, the agreement of the combined linear-cnoidal approach of \cite{svendsen1977wave}
already gave good agreement with experiments while being much less expensive.

In present work we will show that the discrepancy between
conservation of wave height and conservation of energy flux can be resolved
in the context of the KdV theory by using a definition
of the energy flux consistent with the KdV equation \cite{AK4,IK2018},
and without using the fully nonlinear approach of \cite{Dixon}.
Recall that the KdV equation is given in the form
\begin{equation}
    \eta_t + c_0\eta_x + \frac{3}{2}\frac{c_0}{h_0}\eta \eta_x + \frac{c_0 h_0^2}{6}\eta_{xxx} = 0,
    \label{KdVEquation}
\end{equation}
where $\eta(x,t)$ is the deflection of the free surface from rest at a point $x$ and a time $t$,
$g$ is the gravitational acceleration, $h_0$ is the local water depth, and $c_0 = \sqrt{g h_0}$
is the long-wave speed.

Shoaling in the context of steady solutions of the KdV equation is then described as follows.
Given a steady wavetrain of wave height $H$, wavelength $\lambda$ and period $T$,
the variation of the wave as it propagates from depth $h^A$ to $h^B$ is obtained
by imposing conservation of $T$, conservation of the energy flux $q_E$, and
balance of forces using the bottom forcing by the pressure force, and the radiation stress.

In order to execute this plan, one needs to have in hand formulations for the energy flux
$q_E$ and the radiation stress $S_{xx}$ in the context of the KdV equation.
These expressions can be developed using ideas first put forward in \cite{AK2,AK4}.
Using the methods in these papers, it can be seen that we have
\begin{equation*}
    q_E = c_0^3
    \Big{(}
    \frac{1}{h_0}\eta^2 + \frac{5}{4h_0^2} + \frac{h_0}{2}\eta \eta_{xx}
    \Big{)}
\end{equation*}
and
\begin{equation*}
\overline{S}_{xx} =           \rho g  \Big{(}
           \overline{\eta} 
           +
           \frac{3}{2}\overline{\eta^2} 
           +
           \frac{h_0^3}{3}\overline{\eta}_{xx}\Big{)}.
\end{equation*}
Note that the energy flux is defined in a pointwise sense while the radiation stress
is defined in an average sense. However, for the shoaling problem $q_E$ will also
be averaged over one wave period.

The plan of the paper is as follows.
In the next section, the formulation of momentum and energy balance
in the context of the KdV equation will be recalled, leading to
the expression for flow force $q_I$ and energy flux $q_E$.
In Section 3, a formulation of the radiation stress consistent
with the KdV equation will be found.
Section 4 contains the formulation of the nonlinear shoaling equations,
details on the numerical implementation and a comparison with wave tank
experiments with particular focus on the set-down.
Section 6 details the comparison with other nonlinear shoaling theories
without set-down, and in particular with the shoaling curves obtained by 
Svendsen and Buhr Hansen \cite{svendsen1977wave}.
In Section 7, we provide a comparison with the experimental results
of \cite{svendsen1977wave}. In the Conclusion we put our work into
context and mention possibilities for further work.

\section{Momentum and energy balance in the KdV approximation}
%

\begin{figure}
  \includegraphics[scale=0.4]{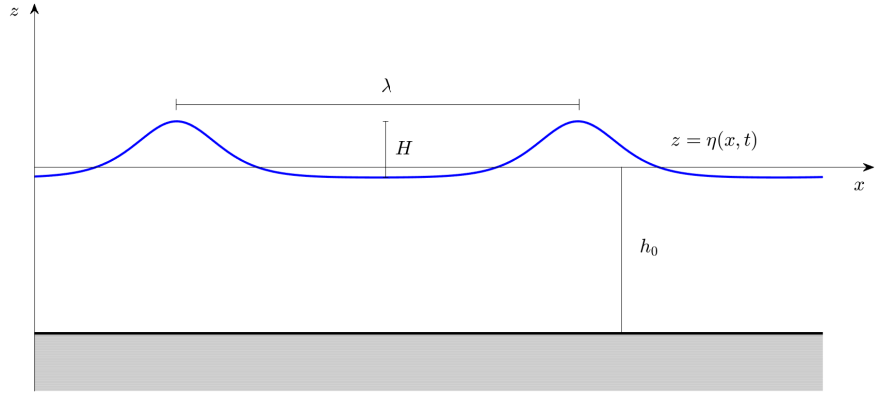}
  \centering
  \caption{\small A periodic wave of wave height $H$ and wavelength $\lambda$ propagating 
                  at the free surface of a fluid of undisturbed depth $h_0$. The free surface
                  is given by $z=\eta(x,t)$.}
  \label{Schematic}
\end{figure}

We start this section with a brief description of the water-wave problem
of an inviscid, incompressible and homogeneous fluid with a free surface.
Due to the assumption that the relative change in water depth
over one wavelength is less than the wave steepness, the problem can be 
locally formulated with a constant undisturbed depth $h_0$ (see Figure \ref{Schematic}).%
%
%
If the flow is assumed to be irrotational, the problem can be formulated
in terms of a velocity potential $\phi(x,z,t)$ in addition to
the surface deflection $\eta(x,t)$.
The velocity field $(u(x,z,t),v(x,z,t))$ is then given as the gradient of $\phi$,
and the pressure $P$ can be found using the Bernoulli equation.
In terms of $\phi$ and $\eta$,
the problem is described by the Laplace equation
\begin{equation}
    \hspace{3.5cm} \phi_{xx} + \phi_{zz} = 0 \hspace{0.5cm} \text{for} \hspace{0.2cm} -h_0 <z < \eta(x,t)
\end{equation}
in the fluid domain, and the boundary conditions
\begin{align*}
    \eta_t + \phi_x\eta_x - \phi_z & = 0 \hspace{0.5cm} \text{on} \hspace{0.2cm} z = \eta(x,t),\\ 
    \phi_t + \frac{1}{2}(\phi_x^2 + \phi_z^2) + g\eta & = 0 \hspace{0.5cm} \text{on} \hspace{0.2cm} z = \eta(x,t), \\
    \phi_z & = 0 \hspace{0.5cm} \text{on} \hspace{0.2cm} z = -h_0.
\end{align*}

It is well known that this problem is difficult to treat both mathematically and numerically. 
In particular, it is not known whether solutions exist on relevant time scales \cite{LannesBOOK},
and numerical discretization of the full problem on the field scale remains completely out of
reach. 
Thus in practical situations, an asymptotic approximation of the Euler equations is usually required. 
The traditional asymptotic theories are the linearized (Airy) theory
and the hyperbolic (Saint-Venant) shallow-water theory such as outlined in \cite{Stoker}.
The work of Boussinesq \cite{Bouss} and Korteweg and de Vries \cite{KdV} led to another
asymptotic regime, the so-called Boussinesq scaling which can be used for long waves of small to moderate
amplitude. In the present work, we consider waves in the Boussinesq regime, and since we
consider waves propagating in the shoreward direction only, 
we restrict attention to the Korteweg-de Vries (KdV) equation.
Indeed, the KdV equation describes unidirectional waves in the presence of weakly nonlinear and dispersive effects
in the case when these corrections are approximately balanced. 
In order to make this statement precise, it is useful to define the relative amplitude $\alpha = a/h_0$ 
and the shallowness parameter $\beta = h_0^2/\lambda^2$, where $a=H/2$ denotes a 
representative amplitude and $\lambda$ a representative wavelength of the wavefield to be studied. 
Using an asymptotic expansion and a dimensional argument it can be shown that
the KdV equation is a good model for long waves of small amplitude
if terms of order $\mathcal{O}(\alpha^2,\alpha \beta, \beta^2)$ are neglected.

In order to derive the KdV equation, 
one defines the non-dimensional variables $x = \lambda \tilde{x}$, 
$z = h_0(\tilde{z}-1)$,  $t = \frac{\lambda}{c_0}\tilde{t}$,  
and $\phi = \frac{ga\lambda}{c_0} \tilde{\phi}$, 
then assumes that the velocity potential takes the form 
\begin{align}
    \tilde{\phi} = \tilde{f} - \beta \frac{\tilde{z}^2}{2}\tilde{f}_{\tilde{x}\tilde{x}} + \mathcal{O}(\beta^2),
    \label{non dim potential}
\end{align}
where $\tilde{f}$ is the velocity potential evaluated at the bed.
Following the description in \cite{Whitham}, 
the free-surface boundary conditions can then be used to obtain the KdV equation \eqref{KdVEquation}.
In the process, it comes to light that
the horizontal velocity at the bottom for a right-moving wave is given by
\begin{equation}
\tilde{u} = \tilde{\eta} - \frac{1}{4} \alpha \tilde{\eta}^2 
+ \frac{1}{3}\beta \tilde{\eta}_{\tilde{x}\tilde{x}\tilde{x}} + \mathcal{O}(\alpha^2 + \beta^2).
    \label{horizontal velocity}
\end{equation}

Moreover, as noted for example in \cite{AK4},
combining $\eqref{horizontal velocity}$ and the derivative of $\eqref{non dim potential}$ 
with respect to $x$ one finds that the horizontal component of the velocity field 
is given by
\begin{equation}
\tilde{\phi}_{\tilde{x}}(\tilde{x},\tilde{z},\tilde{t}) 
= \tilde{\eta} - \frac{1}{4} \alpha \tilde{\eta}^2 
+ \beta \Big( \frac{1}{3}-\frac{\tilde{z}^2}{2} \Big) \tilde{\eta}_{\tilde{x}\tilde{x}} 
+ \mathcal{O}(\alpha^2,\alpha \beta, \beta^2)
    \label{non dim potentialX}
\end{equation}
in the KdV approximation. This relation will be needed later 
in the derivation of the the momentum and energy balance, 
and the formulation of the radiation stress. 
In addition, we need the vertical component of the velocity field and the pressure.
The vertical velocity is given by
\begin{align*}
\tilde{\phi}_{\tilde{z}}(\tilde{x},\tilde{z},\tilde{t})  
= - \beta \tilde{z} \tilde{\eta}_{\tilde{x}} + \mathcal{O}(\alpha \beta, \beta^2).
\end{align*}
In order to express the pressure in terms of the surface excursion $\eta$, 
assuming unit density for the moment, 
one first defines the dynamic pressure in the usual way:
\begin{equation*}
    P' = P - P_{\text{atm}} + gz, \quad P' = ag \tilde{P}'.
\end{equation*}
Then using the Bernoulli equation and following the computations
outlined in \cite{AK4} leads to the non-dimensional perturbation pressure 
in the KdV approximation: 
\begin{equation}
    \tilde{P}' = \tilde{\eta} - \frac{1}{2}\beta (\tilde{z}^2 - 1)\tilde{\eta}_{\tilde{x}\tilde{x}} + \mathcal{O}(\alpha\beta, \beta^2).
\end{equation}
%
Approximations of the momentum and energy balance laws in the KdV scaling regime
will now be derived. In the context of the full Euler equations with surface boundary conditions, 
the momentum balance is expressed by
\begin{equation}
\frac{\partial }{\partial t} 
\int_{-h_0}^{\eta} \phi_x dz + \frac{\partial}{\partial x} \int_{-h_0}^{\eta} \big\{ \phi_x^2 + P \big\} dz = 0.
\label{momentum balance law}
\end{equation}
Using the previously defined scaling, the momentum balance is
\begin{equation*}
\alpha \frac{\partial}{\partial \tilde{t}} \int_{0}^{1+\alpha \tilde{\eta}}\tilde{\phi}_{\tilde{x}} h_0 d\tilde{z}
+ \frac{\partial}{\partial \tilde{x}} 
\int_{0}^{1+\alpha \tilde{\eta}} \big\{ \alpha^2 \tilde{\phi}^2_{\tilde{x}} + \alpha \tilde{P}' +  (1-\tilde{z})   
\big\} h_0 d\tilde{z} = 0.
\end{equation*}
Substituting the two expressions $\tilde{\phi}_x$, $\tilde{P}'$ and evaluating the integral one finds 
\begin{equation*}
     \Big( \alpha\tilde{\eta} + \frac{3}{4}\alpha^2 \tilde{\eta}^2 
                              + \frac{1}{6} \alpha\beta \tilde{\eta}_{\tilde{x}\tilde{x}} \Big)_{\tilde{t}}
    + \Big( \frac{1}{2} + \alpha \tilde{\eta} + \frac{3}{2}\alpha \tilde{\eta}^2 
                        + \frac{1}{3}\beta \tilde{\eta}_{\tilde{x}\tilde{x}} \Big)_{\tilde{x}} 
    = \mathcal{O}(\alpha^2,\alpha\beta,\beta^2).
\end{equation*}
From this relation we identify the non-dimensional momentum density 
\begin{equation*}
    \tilde{I} = \alpha\tilde{\eta} + \frac{3}{4}\alpha^2 \tilde{\eta}^2 + \frac{1}{6} \alpha\beta \tilde{\eta}_{\tilde{x}\tilde{x}},
\end{equation*}
and the non-dimensional momentum flux 
\begin{equation*}
    \tilde{q_I} = \frac{1}{2} + \alpha \tilde{\eta} + \frac{3}{2}\alpha \tilde{\eta}^2 
                             + \frac{1}{3}\beta \tilde{\eta}_{\tilde{x}\tilde{x}}.
\end{equation*}
Returning the expression to its dimensional forms through the scaling
$I = c_0h_0 \tilde{I}$ and $q_I = c_0^2h_0\tilde{q}_I$ yields
\begin{equation*}
    I = c_0\Big{(}\eta + \frac{3}{4h_0}\eta^2 + \frac{h_0^2}{6}\eta_{xx}\Big{)},
\end{equation*}
and
\begin{equation}
    q_I = c_0^2\Big{(}\frac{h_0}{2} + \eta + \frac{3}{2h_0}\eta^2 + \frac{h_0^2}{3}\eta_{xx}\Big{)}.
    \label{momentum flux KdV}
\end{equation}
Note that the expression for $q_I$ combines the momentum flux and the pressure force.
Following Benjamin and Lighthill \cite{BL}, we will use the term {\it flow force} for this quantity.
Similarly, we can give the energy balance in the full Euler equations as
\begin{equation*}
    \frac{\partial}{\partial t} \int_{-h_0}^{\eta} \Big{\{} \frac{1}{2} |\nabla \phi|^2 + g(z + h_0) \Big{\}} dz + \frac{\partial}{\partial x} \int_{-h_0}^{\eta} \Big{\{} \frac{1}{2} |\nabla \phi |^2 + g(z+h_0) + P \Big{\}} \phi_x dz =0,
\end{equation*}
and following the same procedure as above will lead to the expressions
\begin{equation*}
    E = c_0^2
    \Big(
    \frac{1}{h_0}\eta^2 + \frac{1}{4h_0^2}\eta^3 + \frac{h_0}{6} \eta\eta_{xx} + \frac{h_0}{6} \eta_{x}^2
    \Big),
\end{equation*}
and 
\begin{equation}
    q_E = c_0^3
    \Big(
    \frac{1}{h_0}\eta^2 + \frac{5}{4h_0^2}\eta^3 + \frac{h_0}{2}\eta \eta_{xx}
    \Big),
    \label{EnergyFlux}
\end{equation}
for the energy density and energy flux respectively.

Note that $q_E$ is of second order, while the energy flux in the linear approximation
is of first order.
Indeed, it will be instructive to compare \eqref{EnergyFlux} to the well known expression 
for the energy flux in the linear theory.
Denoting the amplitude of a linear wave by $a=H/2$ and the group velocity by $C_g$ as before,
it can shown (see for example \cite{kundu2008fluid}) that
the energy flux averaged over one period is
$$
\overline{q_E^{\mathrm{linear}}} = g \overline{\eta^2} C_g = \frac{1}{2} g a^2 C_g.
$$
On the other hand, for cnoidal waves of very large wavelength and very small amplitude 
\eqref{EnergyFlux} can be approximated by
$$
q_{E} \sim c_0^3 \frac{1}{h_0}\eta^2 = c_0 g \eta^2.
$$
Averaging over one period and using the fact that the cnoidal wave is similar to a 
sinusoidal function for very small amplitudes, we obtain
$$
\overline{q_{E}} \sim \frac{1}{2} g a^2 c_0.
$$
Finally, the two expressions are seen to be approximately equal if it is recognized 
that $c_0$ is the group velocity for long linear waves in shallow water.

Note that in previous works on cnoidal shoaling such as \cite{svendsen1972,svendsen1977wave}, 
the linear formulation of the energy flux was used. In addition to being obviously inconsistent,
this approach also had practical drawbacks. Indeed as already mentioned, the method put forward
in \cite{svendsen1977wave,svendsen1972} led to a discontinuity in the wave height at the matching
point between linear and nonlinear theory.
As will be shown momentarily, keeping the nonlinear terms in the expression
for $q_E$ resolves this problem, and indeed is essential in order to obtain 
the correct shoaling behavior for larger wave amplitudes.

In the next section, we will use the definition of the flow force $q_I$, and ideas outlined above 
to derive an expression for the radiation stress in the context of the KdV equation. 
Incorporating the radiation stress into the shoaling equations will then enable us to identify
the set-down in addition to the wave height of a shoaling wave.

\section{Radiation stress in the KdV approximation}

In linear water-wave theory, the radiation stress is well understood as a second-order response
to a periodic wave train which is akin to the energy flux. Wave radiation stress is an important ingredient in
the study of wave-current interactions, and also plays a major role in the understanding of wave setup
in the context of breaking waves. The reader may consult \cite{longuet1962radiation} for the mathematical
development of an expression for the radiation stress in the context of linear waves. 
A discussion of the radiation stress based on a more heuristic physical understanding is given by
Longuet-Higgins and Stewart \cite{longuet1964radiation}.
In the present work, we aim to expand the definition of the radiation stress to the nonlinear case,
in particular in the context of the KdV equation.
Using the momentum balance equation derived in the previous section, 
a nonlinear version of the radiation stress will be formulated.
With this expression in hand,
some consequences of the radiation stress on wave shoaling will be investigated.

Analyzing the definition given in \cite{longuet1964radiation},
one can simply think of radiation stress as the total flow force 
of a progressive wave averaged over one period minus the hydrostatic pressure force at rest. 
As previously discussed we consider 
a wave propagating solely in the $x-$direction, and neglect all transverse effects. 
Then the definition of the principal component of the radiation stress 
is given by
\begin{equation*}
   \overline{S}_{xx} = \overline{\int_{-h_0}^{\eta} (\rho u^2 + P ) dz} + \int_{-h_0}^{0}\rho g z dz.
\end{equation*}

\noindent
The first term expresses the total flux of momentum across a plane integrated 
from the bottom to the free surface and with unit width, and as usual, 
the overbar denotes that the average over one wave period with respect to time is taken.
The second term expresses the flow force in the absence of any wave motion.

In the context of the KdV equation, the flow force is given 
by equation \eqref{momentum flux KdV} when rescaled for a fluid with unit density.
Consequently, the $x-$component of the radiation stress is obtained in the
KdV theory as
\begin{align}
   \overline{S}_{xx}  & =  \overline{q_I}  + \int_{-h_0}^0 \rho g z dz,\notag  \\
& = 
           \rho g h_0 \Big{(}\frac{h_0}{2} 
           + 
           h_0 \overline{\eta} 
           +
           \frac{3}{2h_0}\overline{\eta^2} 
           +
           \frac{h_0^2}{3}\overline{\eta}_{xx}\Big{)} 
           -
            \frac{1}{2} \rho g h_0^2,\notag  \\
& = 
           \rho g  \Big{(}
           h_0 \overline{\eta} 
           +
           \frac{3}{2}\overline{\eta^2} 
           +
           \frac{h_0^3}{3}\overline{\eta}_{xx}\Big{)}.
           \label{RadiationStressXX}
\end{align}
%
%
%

\begin{figure}
  \includegraphics[scale=0.4]{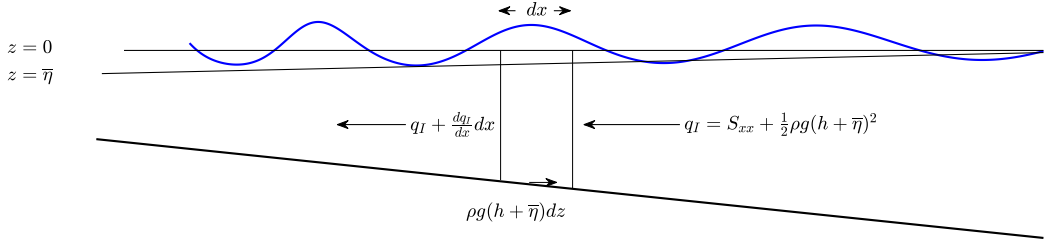}
  \centering
  \caption{\small Schematic of the forces acting on a control volume given by a differential
                  interval of length $dx$ and reaching through the whole fluid column.}
  \label{Balance of forces}
\end{figure}

This formulation represents an improvement over the formulation given in \cite{svendsen2006introduction}
which only recorded the middle term $ \rho g \frac{3}{2} \overline{\eta^2}$ in the final expression for the
radiation stress. A preliminary version of this quantity was also presented in \cite{khorsand2017flow}.
In order to apply the radiation stress to the formulation of the shoaling problem,
let us consider a wave encountering a gently sloping beach. 
As indicated in Figure \ref{Balance of forces},
the momentum flux is reduced in the onshore direction due to an opposing horizontal force 
exerted by the bed, and opposing the fluid pressure.
%
%
%
When accounting for the set-down of the mean surface level,
the flow force on the offshore side of
the boundary of the control volume 
is given in terms of radiation stress by
\begin{equation*}
q_I =  S_{xx} + \int_{-h}^{\overline{\eta}} \rho g (\overline{\eta} - z) dz = S_{xx} + \frac{1}{2} \rho g (\overline{\eta} + h)^2.
\end{equation*}
This relation can be found for example in \cite{dean1984water, longuet1964radiation}. 
As already mentioned above, the momentum flux is reduced by the reaction force exerted by the
sea bed on the fluid due to the weight of the fluid. This force is equal in magnitude to the
horizontal component of the pressure force at the bottom
$\rho g (h + \overline{\eta}) dz$, where $dz$ denotes the vertical variation in the sea bed
over the horizontal distance $dx$.
Thus the horizontal component of the pressure force exerted on the bottom can be written as
\begin{equation}
    \rho g (h + \overline{\eta}) dz = \rho g (h + \overline{\eta}) h_x dx.
    \label{set-down pressure bottom}
\end{equation}
\noindent
As indicated in Figure \ref{Balance of forces}, the difference in flow force 
between the shoreward face and the offshore face
of the control volume is given by
 \begin{equation}
     \frac{d}{dx}\Big[ S_{xx} + \frac{1}{2}\rho g (h + \overline{\eta})^2 \Big] dx.
     \label{ingreace of flux}
 \end{equation}
\noindent
Evidently, momentum balance requires \eqref{set-down pressure bottom} and \eqref{ingreace of flux} 
to be equal, and simplifying yields the relation
\begin{equation}
    \frac{dS_{xx}}{dx} + \rho g (\overline{\eta} + h) \frac{d\overline{\eta}}{dx} = 0.
    \label{momentum balance}
\end{equation}
In order to develop the shoaling equation, we integrate over the control interval
$[x, x+dx]$ to get
\begin{equation*}
\int_{x}^{x+dx}\frac{dS_{xx}}{dx} + \int_{x}^{x+dx} \rho g \overline{\eta} \frac{d \overline{\eta} }{dx} 
                              + \int_{x}^{x+dx} \rho g h \frac{d\overline{\eta} }{dx}=0.
\end{equation*}
%
%
The first two integrals are straightforward to compute.
The third integral is evaluated by using integration by parts and then approximated using the trapezoidal rule. 
Defining the difference of any quantity $F$ as $\Delta F = F|_{x+dx} -  F|_{x}$ we get, the relation
\begin{equation}
    \Delta S_{xx} = - \frac{\rho g}{2}\Delta \overline{\eta}^2 - \frac{\rho g}{2}(2h+h_x dx)\Delta \overline{\eta}.
\end{equation}

\noindent
Recall that we are interested in the averaged behavior of the wave height for a wave-train approaching the beach. 
We therefore  find it useful to define the changes in radiation stress averaged over a period by
\begin{equation}
   \Delta \overline{S}_{xx}   = -\frac{1}{T}\int_0^{T} \Big{\{} 
    \frac{\rho g}{2}\Delta \overline{\eta}^2 + \frac{\rho g}{2}(h_0+h)\Delta \overline{\eta}
   \Big{\}} dt,
   \label{RadiationStressEquation}
\end{equation}
\noindent
where $\overline{S}_{xx}$ is defined by \eqref{RadiationStressXX}. 
With this identity in hand, it will be possible to include the development of the set-down $\overline{\eta}$
in the shoaling problem. The formulation of the complete shoaling equation will be the subject of the next section.

\section{The nonlinear shoaling equations}
\label{ShoalingModel}
%
%
An explicit set of equations describing the shoaling of long waves on a gently 
sloping beach will now be given. The idea is to use the well known cnoidal wave solution 
for periodic waves in the KdV equation together with the assumption that the bottom slope
is gentle enough so that reflections can be neglected, and the waves are able to adjust
adiabatically to the new local depth. The wave profile $\eta$ will distort, but
the underlying cnoidal shape and periodicity are preserved.

%
A solution of the KdV equation for periodic waves of constant form was first 
discovered by Korteweg and de Vries in 1895 \cite{KdV}.
Assuming a traveling-wave solution of constant shape, one may make the ansatz $\eta (x,t) = f(x-ct)$,
and reduce the KdV equation to an ordinary differential equation of the form
\begin{equation}
    -\frac{h_0^2}{3}(f')^2 = F(f) =  \frac{1}{4} \frac{c_0}{h_0}f^3 
                                    + \Big( \frac{c-c_0}{2}\Big) f^2 + Af + B, 
    \label{ODE-Poly}
\end{equation}
with $A$ and $B$ being constants of integration. 
This differential equation has a number of solutions, but for practical applications 
only periodic real-valued and bounded solutions correspond to realistic wave profiles.
Since $F(f)$ is a third-order polynomial it has three roots.
Periodic solutions occur only in the case where the roots are distinct,
so that they can be labeled $f_3<f_2<f_1$,
and this convention allows us to write
\begin{equation*}
 F(f) = (f - f_1) (f - f_2) (f - f_3).
\end{equation*}
\vspace{-0.5cm}
\begin{figure}[H]
  \includegraphics[scale=0.48]{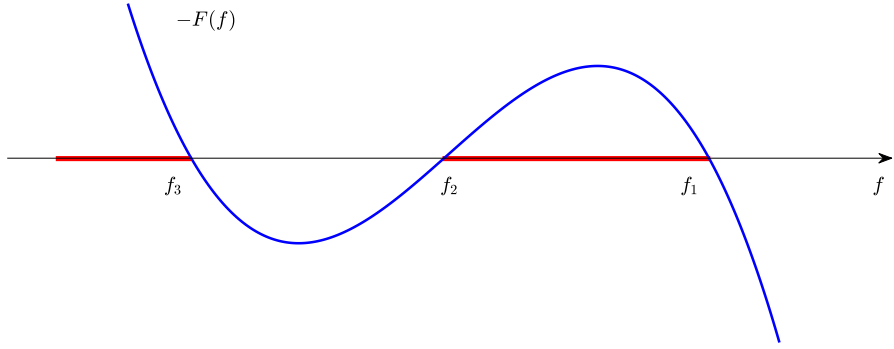}
  \centering
  \vspace{-0.5cm}
  \caption{\small Graph of the function $-F(f)$. The roots are labelled by $f_3 < f_2 <f_1$.
                 The red lines denote the values of $f$ such that $-F\geq 0$.}
  \label{RootsoftheStdKdV}
\end{figure}
\noindent
Requiring the solution to be real, it can be seen from \eqref{ODE-Poly} that 
$-F(f)$ must be positive.
Examining the phase plane plot in Figure \ref{RootsoftheStdKdV}, it is clear
that periodic solutions will oscillate between $f_2$ and $f_1$. 
Since $f_2 < f_1$ by assumption,  $f_1$ will denote the wavecrest while $f_2$ will be the trough.
Consequently, the wave height $H$ is given by the difference $H = f_1 - f_2$. 
As is well known, the periodic solutions of \eqref{ODE-Poly}
are given in terms of the Jacobian elliptic function $cn$. 
The solution of the KdV equation is then
\begin{equation}
    \eta(x,t) =
    f_2 - (f_2 - f_1)
    \text{cn}^2 \Big( 2K(m) \big( {\textstyle \frac{t}{T}} - {\textstyle \frac{x}{\lambda}} \big), m \Big),
    \label{CnoidalWaveSolution}
\end{equation}
%
%
where $m$ is the elliptic modulus defined by $m=\frac{f_{1}-f_{2}}{f_{1}-f_{3}}$,
and $K(m)$ is the complete elliptic integral of first kind \cite{lawden2013elliptic}.
The wavespeed $c$ and the wavelength $\lambda$ are then given by
\begin{equation}
    \hspace{1cm} c = c_0 \Big{(} 1 + \frac{f_1 + f_2 + f_3}{2h_0}\Big{)},
    \hspace{1cm} \lambda = K(m) \sqrt{\frac{16 h_0^3}{3(f_1 - f_3)}}.
    \label{parameters}
\end{equation}
In order to use these formulas in practice, it is convenient to 
take the wave height $H$, the mean surface level $\overline{\eta}$
and the elliptic parameter $m$ as given parameters.
The roots of $F$ can then by computed explicitly and are given as follows.
\begin{equation}
\begin{dcases}
    f_3 & =  \overline{\eta} - \frac{H E(m)}{mK(m)},
    \\
    f_1 & = f_3 + \frac{H}{m},
    \\
    f_2 & = f_1 - H.
    \label{roots}
\end{dcases}
\end{equation}
%
%
The shoaling problem can be formulated as follows. Suppose we know that a periodic wave train
of wave height $H$, wavelength $\lambda$ and mean-surface level $\overline{\eta}$
is given at a local depth $h^A$. Assuming that the waves can be approximately described by the
the KdV equation, a unique cnoidal wave solution can be found, and
the frequency $\nu$, the energy flux $q_E$ and the radiation stress $S_{xx}$ be found.
The shoaling problem now consists of finding an appropriate cnoidal solution
at a smaller depth $h^B$.
Relying on the conservation of frequency and energy flux, 
and a prescribed change in the radiation stress, the shoaling equations can be written as
\begin{align}
         \nu^B(f_1,f_2,f_3,h^B) \  = & \ \ \nu^A \ = \ \frac{c^A}{\lambda^A},
        \label{SystemOne}
        \\
         \overline{q}_E^B(f_1,f_2,f_3,h^B)  \  = & \ \ \overline{q}_E^A  
                                           \ = \  \frac{1}{T}\int_0^{T} \Big\{\frac{1}{h^A} \eta^2 
                                                    + \frac{5}{4 (h^A)^2}\eta^3 
                                                    +  \frac{h^A}{2}\eta \eta_{xx} \Big\} \hspace{1mm} dt,
        \label{SystemTwo}
        \\ 
\overline{S}^B_{xx}(f_1,f_2,f_3,h^B) \ =  &  \ \ \overline{S}^A_{xx}  -\frac{1}{T}\int_0^{T} \Big\{ 
            \frac{\rho g}{2}\Delta\eta^2 + \frac{\rho g}{2}(h^B+h^A)\Delta\eta
            \Big\} dt.
   \label{SystemThree}
\end{align}

\noindent
The first equation is conservation of frequency $\nu$, 
while the second equation expresses conservation of energy flux integrated over a period $T$ \cite{AK4}. 
Finally, the third equation is \eqref{RadiationStressEquation}, 
and is indeed a formulation of conservation of momentum
expressed in terms of radiation stress. 
In previous work \cite{khorsand2014shoaling}, a similar system was found,
and then solved for the parameters $f_1$, $f_2$ and $f_3$.
However, the method used in \cite{khorsand2014shoaling} had several drawbacks.
First, due to a lack of a definition of radiation stress, the equation for
momentum conservation was replaced by conservation of the mean fluid level.
This approach was therefore not able to predict the wave set-down.
Moreover, solving for the parameters  $f_1$, $f_2$ and $f_3$ was not
optimal from a numerical point of view, and the algorithm terminated long
before the highest wave was reached.
In the present work, we use the explicit form of the parameters to solve
directly for $H$, $m$ and $\overline{\eta}$.
This approach is more expedient from a numerical point of view,
and allows us to go much higher up on the shoaling curve.
In terms of the wave height $H$, the conservation of frequency
turns into the following cubic equation for $H$:
\begin{equation}
    \frac{-3g(\frac{3E(m)}{mK(m)} - \frac{2}{m} + 1)^2}{64K(m)^2h^4m}H^3 
    +
    \frac{3g(\frac{3\overline{\eta}}{2h} + 1)(\frac{3E(m)}{mK(m)} - \frac{2}{m} + 1)}{16K(m)^2h^3m}H^2 
    +
    -\frac{3g(\frac{3\overline{\eta}}{2h} + 1)^2}{16K(m)^2h^2m}H 
    +
    \frac{1}{T^2} = 0.
    \label{WaveHeight}
\end{equation}
%
%
%
Similarly we can manipulate the equation describing the changes in radiation stress 
to find an expression for the set-down. Indeed, \eqref{SystemThree} reduces to the quadratic equation
\begin{align}
- \overline{\eta}^2  + \mathcal{A}
\overline{\eta} + \mathcal{B} = 0,
    \label{setDown}
\end{align}
\noindent
with
%
\begin{align*}
\mathcal{A} = \Big( 3H - \frac{3h^B}{2} + \frac{h^A}{2} - 3H\overline{\text{cn}^2}(m)
                                        - \frac{3H}{m} + \frac{3HE(m)}{mK(m)} \Big)
\end{align*}
%
and
%
%
\begin{align*}
\mathcal{B} =
\overline{S}^A_{xx}(m)
-
\frac{\overline{\eta^2}^A(m)}{2}
-
\frac{3}{2}H^2\overline{\text{cn}^4}(m)
 -
\frac{(h^B)^3 \overline{\eta_{xx}}^B(m)}{3}
+\frac{3}{2}h^A \overline{\eta}^A(m)-
\frac{1}{2}h^B\overline{\eta}^A(m) \\
-
\frac{3}{2}( H - \frac{H}{m} + \frac{HE(m)}{mK(m)})^2
+
3H \overline{\text{cn}^2}(m)(H -
\frac{H}{m} +
\frac{HE(m}{mK(m)} ).
\end{align*}

Note that we had to integrate various powers and derivatives of $\text{cn}^2({\xi;m})$. 
These formulas are based on calculations that can be found in \cite{lawden2013elliptic}
and \cite{abramowitz1965handbook}, and are given in explicit form in the appendix. 
Having this representation in hand, we can in principle write \eqref{WaveHeight} 
as $H = F(m,\overline{\eta})$ and take \eqref{setDown} 
to be $\overline{\eta} = G(m,H)$ in explicit terms as two coupled equations. 
To solve these two equations simultaneously, we may iterate between 
them at current local depth $h^B$ as follows; 
first initialize the procedure with $\overline{\eta}^0(m)$ and then find $H$ by
\begin{equation*}
    H(m)^{i+1} = F(m,\overline{\eta}^{i}(m)),
\end{equation*}
%
solving the cubic equation. This can in turn be used to update the set-down by
\begin{equation*}
    \overline{\eta}(m)^{i+1} = G(m, H^{i+1}(m)).
\end{equation*}
%
%
Repeating this process, we continue to approximate $H$ and $\overline{\eta}$ 
until a stopping criterion has been reached. 
Using this reduction allows us to solve the nonlinear system 
\eqref{SystemOne},\eqref{SystemTwo},\eqref{SystemThree},
by the following procedure. From the value of $m$ at depth $h^A$,
we increment up, at each step iterating the above equations for
$H$ and $\overline{\eta}$, and then checking whether \eqref{SystemTwo}
is satisfied to a specified tolerance. If that is the case,
we consider the system solved at depth $h^B$, and move on
to the next step with depth $h^C$, thereby moving up the slope.

\section{Implementation of the shoaling equations}
\label{Implementation}
Having explained the nonlinear shoaling equations, and the strategy for finding approximate solutions,
we will now implement the equations and produce shoaling curves for various deep-water data. 
The method used here proceeds in three stages, 
such as originally developed in \cite{svendsen1972, svendsen1977wave}. 
Suppose an incoming wave of wavelength $\lambda_0$ is given at a starting
depth $h_0$. While this depth may not be {\em deep} water, it is assumed
that $h_0 > 0.28 \lambda_0$, so that we label this starting value
the deep-water wavelength.
First, the linear shoaling equation is used up to the point $h/\lambda_0 = 0.1$. 
At this point the cnoidal theory is valid \cite{skovgaard1974sinusoidal}, 
and we propose a matching technique to obtain the fundamental parameters 
of the nonlinear wave. Lastly, we use the nonlinear KdV theory 
to follow the shoaling curve into the nonlinear region. \\

\noindent
\textbf{Stage 1. Linear theory:} 
The essential ingredient in this step is the linear dispersion relation
\begin{equation}
    \omega^2 - gk \tanh{(kh)} = 0,
\end{equation}
and the conservation of energy.
Since the circular frequency $\omega$ is conserved, the nonlinear equation
$\omega^2 - gk^B \tanh{(k^Bh^B)} = 0$ needs to be solved numerically for the  wave number $k^B$
using a Newton solver.
The wave height is then found from the conservation of energy as
\begin{equation}
    \label{LinEnergyH}
    H^B = H^A \sqrt{\frac{C_g^A}{C_g^B}},
\end{equation}
where $C_g = \frac{d \omega}{d k}$ is the group velocity depending on the wave
number and depth. The derivation of equation \eqref{LinEnergyH} 
is described in detail by for example in \cite{dean1984water}.
The set-down can be solved similarly using the procedure outlined in \cite{longuet1964radiation}
using the radiation stress. \\

\begin{figure}
  \includegraphics[scale=0.4]{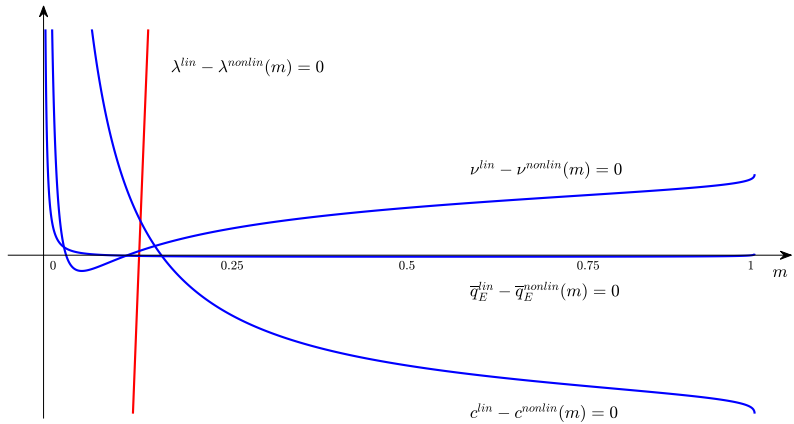}
  \centering
  \caption{\small Root problems defined by the parameters $\lambda, c, \nu, \overline{q}_E$ as functions of $m$.}
  \label{RootsofParameters}
\end{figure}

\noindent
\textbf{Stage 2. Matching linear and nonlinear theory:}  
Due to increasing waveheight, the nonlinear algorithm must be utilized
    on the final stretch of the shoaling curve. The transition must be made in a region
    where both theories are valid, and some general conditions on the transition depth
    are given in \cite{svendsen1977wave}.
In order to initialize the nonlinear solver at the transition depth, three parameters 
need to be specified. We choose to match the wave height $H$ and the set-down
$\overline{\eta}$ with the linear theory. 
In addition one of the parameters $\lambda$, $c$, $\nu$, $\overline{q}_E$ can be be matched.
To do this we choose one of these parameters and then keeping
$H$ and $\overline{\eta}$ constant, 
we  find an elliptic modulus $m\in (0,1)$  which leads to equality in the chosen parameter. 
In principle we can only match one parameter, and it is not clear
which one will be the most convenient.
To investigate this issue, we define the root problems 
\begin{equation*}
    \lambda^{lin}-\lambda^{nonlin}(m) = 0, \hspace{1cm} \hspace{0.5cm} c^{lin} - c^{nonlin}(m) = 0,
\end{equation*}
\begin{equation*}
    \nu^{lin} - \nu^{nonlin}(m) = 0, \hspace{1.4cm} \overline{q}_E^{lin} - \overline{q}_E^{nonlin}(m) = 0.
\end{equation*}
%
%
Here the linear parameter is a fixed constant while the nonlinear quantity 
is a function of $m$ given that $H$ and $\overline{\eta}$ are given. 
Plotting the root problems we observe that the wavelength is most sensitive
to varying $m$, and therefore is the best candidate for a matching parameter. 
Figure \ref{RootsofParameters} shows a particular example. \\

\noindent
\textbf{Stage 3. Cnoidal shoaling:} 
The final step in the shoaling procedure is solving for wave height and set-down using the 
scheme defined in Section 6. First, define $H$ and $\overline{\eta}$ as functions of $m$ 
as given by formula \eqref{WaveHeight} and \eqref{setDown}. 
Then use a nonlinear solver to find $m$ from equation \eqref{SystemTwo}. 
With $m$ in hand, one can determine the wave height $H(m)$ at the current local depth. 
Repeating this procedure will then determine changes of wave height and set-down
at consecutive points, allowing us to move up the slope. 
\begin{figure}[t]
  \includegraphics[scale=0.33]{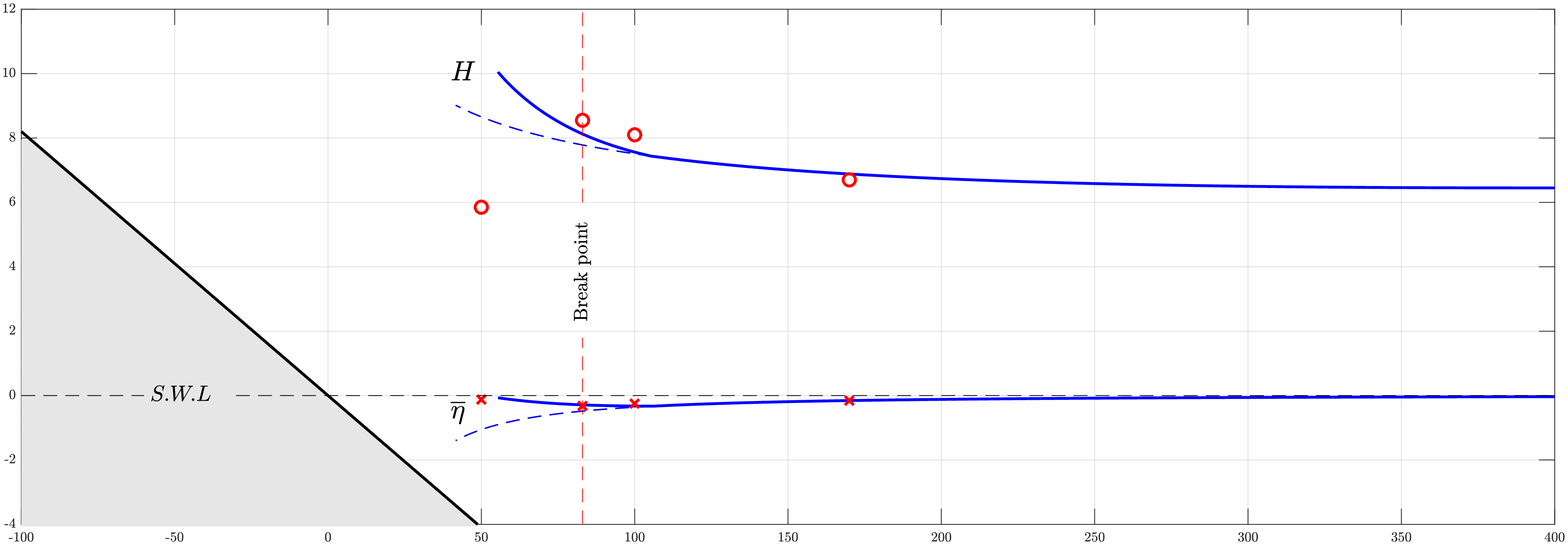}
  \centering
  \includegraphics[scale=0.33]{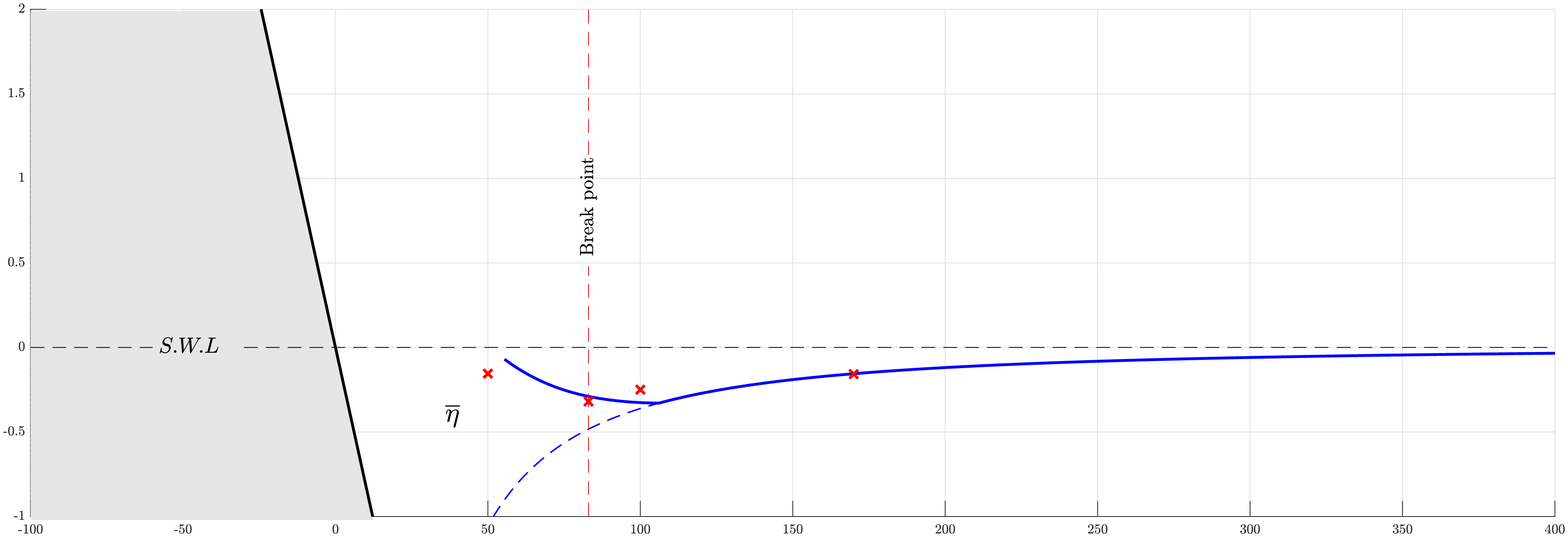}
  \centering
  \caption{\small Profile of the mean water level $\overline{\eta}$ and the wave height $H$ 
                  as functions of the depth, compared to data points taken from $\cite{bowen1968wave}$. 
                  In this case, we have wave period $T=1.14$s incident wave height $H_0 = 6.45$cm and 
                  breaking wave height $H_b = 8.55$cm. In the experiment, the beach slope is 
                  $\mathrm{tan}(\beta) = 0.082$. 
                  The lower panel depicts the same scenario in a different aspect ratio
                  with a zoom-in around the still water line (S.W.L.).}
  \label{BowenShoaling}
\end{figure}

Having defined the shoaling equations and the numerical approach,
we may plot the development of the wave height and set-down 
for a practical shoaling problem. 
A comparison of our numerical results with the wave tank 
data obtained by Bowen et al. \cite{bowen1968wave} 
and the classical linear theory is shown in Figure \ref{BowenShoaling}.
In this figure, the red circles represent the wave height of the shoaling wave
according to the experimental data of \cite{bowen1968wave},
and the red asterics represent the experimentally determined set-down and set-up,
i.e. the deviation of the mean depth from the still water line (S.W.L.).
The dashed blue curves show the prediction of the linear theory 
as already presented in \cite{bowen1968wave}. 
Finally, the solid curves represent the shoaling model 
put forward in the present article. 
The waves under consideration 
had an initial wave length of $\lambda_0 = 202$cm 
and an initial wave height $H_0 = 6.45$cm before they reached the toe of the slope.
The beach had a slope of $1:12$.  

The linear shoaling model agrees well with the experimental data up to moderate wave heights. 
On the other hand, the benefit of the nonlinear formulation can be clearly seen. 
Indeed, the nonlinear theory clearly yields a better fit of both wave height and set-down 
close to the breaking point of the wave. 
One interesting detail in the lower
panel of Figure \ref{BowenShoaling} is that the mean depth appears to rise in the nonlinear
theory while it continues on downwards in the linear theory. 
This {\em nonlinear mid-elevation rise} was acknowledged in \cite{LeMehaute} 
has been found previously with higher-order methods. 
For example, a combination of third-order hyperbolic waves 
near the shore and Stokes waves further offshore was used in \cite{James1974},
and Cokelet's extension of Stokes' approximation of periodic waves was used
in \cite{Stiassnie}.
In both works, a similar rise of the nonlinear set-down was observed.
In the present work, the computed set-down matches the experimental data even well beyond 
the breaking point, but we have to acknowledge that the KdV model ceases to be valid 
for breaking waves.

\section{Shoaling without set-down}
%
%
%
It was argued in \cite{svendsen2006introduction}
that the wave set-down is negligible in many practical cases.
With this assumption, the shoaling problem can be simplified
since only two equations need to be solved.
This approach has been taken in many works on wave shoaling.

The particular aim of this section is to improve the method put forward in \cite{svendsen1972}.
As already mentioned, the use of a first-order approximation of the energy flux
\cite{svendsen1972} gave rise to a discontinuity 
in wave height at the matching point between the linear and their nonlinear theory
due to the inconsistency of the approximation of the energy flux. 
A partial fix was presented in \cite{svendsen1977wave} 
where a continuous shoaling curve is obtained by use of conversion
tables but at the cost of not conserving the energy flux.
More recently, the shoaling theory was extended in \cite{khorsand2014shoaling} 
using the nonlinear definition of the energy flux developed in \cite{AK2,AK4}.
However, due to the specific way the problem was formulated, it was not possible to reach all the way
up the shoaling curve due to numerical instabilities. 

In the present section, it will be shown that with the definition of the shoaling equation
from Section \ref{ShoalingModel}, and the implementation explained in Section \ref{Implementation},
we can find a continuous shoaling curve that reaches all the way up to the highest wave
and agrees well with the more accurate methods of \cite{sakai1980wave} using Cokelet's theory 
and that of \cite{Dixon} using the fully nonlinear steady Euler equations. 

The algorithm is the same as in Section \ref{Implementation},
except that in equation \eqref{setDown} we set $\overline{\eta} = 0$ at each step.
We then use the same three stages as explained in Section \ref{Implementation} 
to determine the development of the wave height in the shoaling region. 
First formulate $H = H(m)$ according to formula \eqref{WaveHeight} given in Section 6. 
Then use the assumption of zero set-down to find the roots given by \eqref{roots}. 
Having the roots as functions of $m$ we may use energy conservation 
to define a nonlinear equation as done in equation \eqref{SystemTwo}. 
Solving for $m$ we are free to determine the wave height at a specified depth $h$. 
\begin{figure}
\centering
\begin{minipage}{.4\textwidth}
  \centering
  \includegraphics[width=1.2\linewidth]{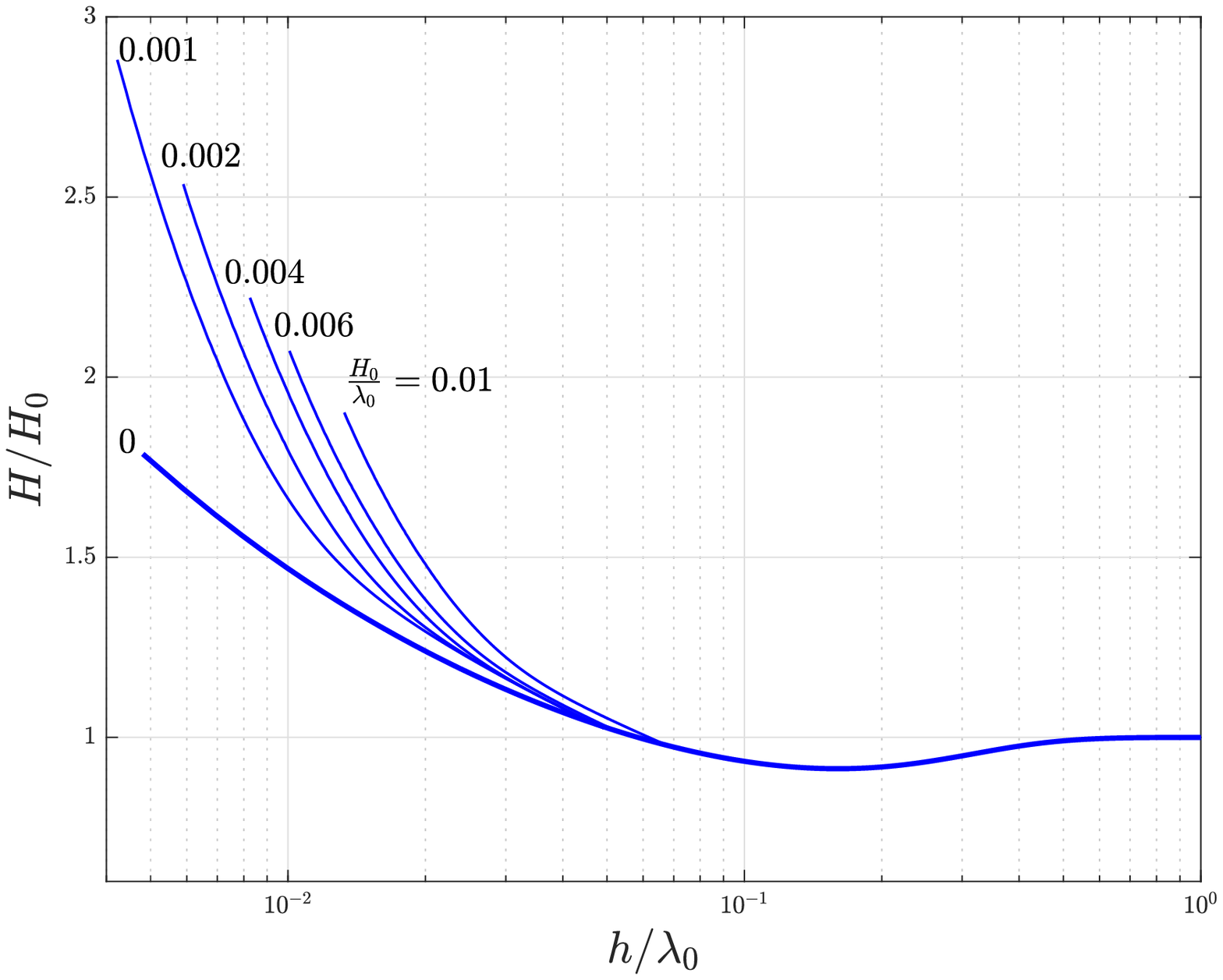}
  \label{SampleHq}
\end{minipage}%
\hspace{1cm}
\begin{minipage}{.4\textwidth}
  \centering
  \includegraphics[width=1.2\linewidth]{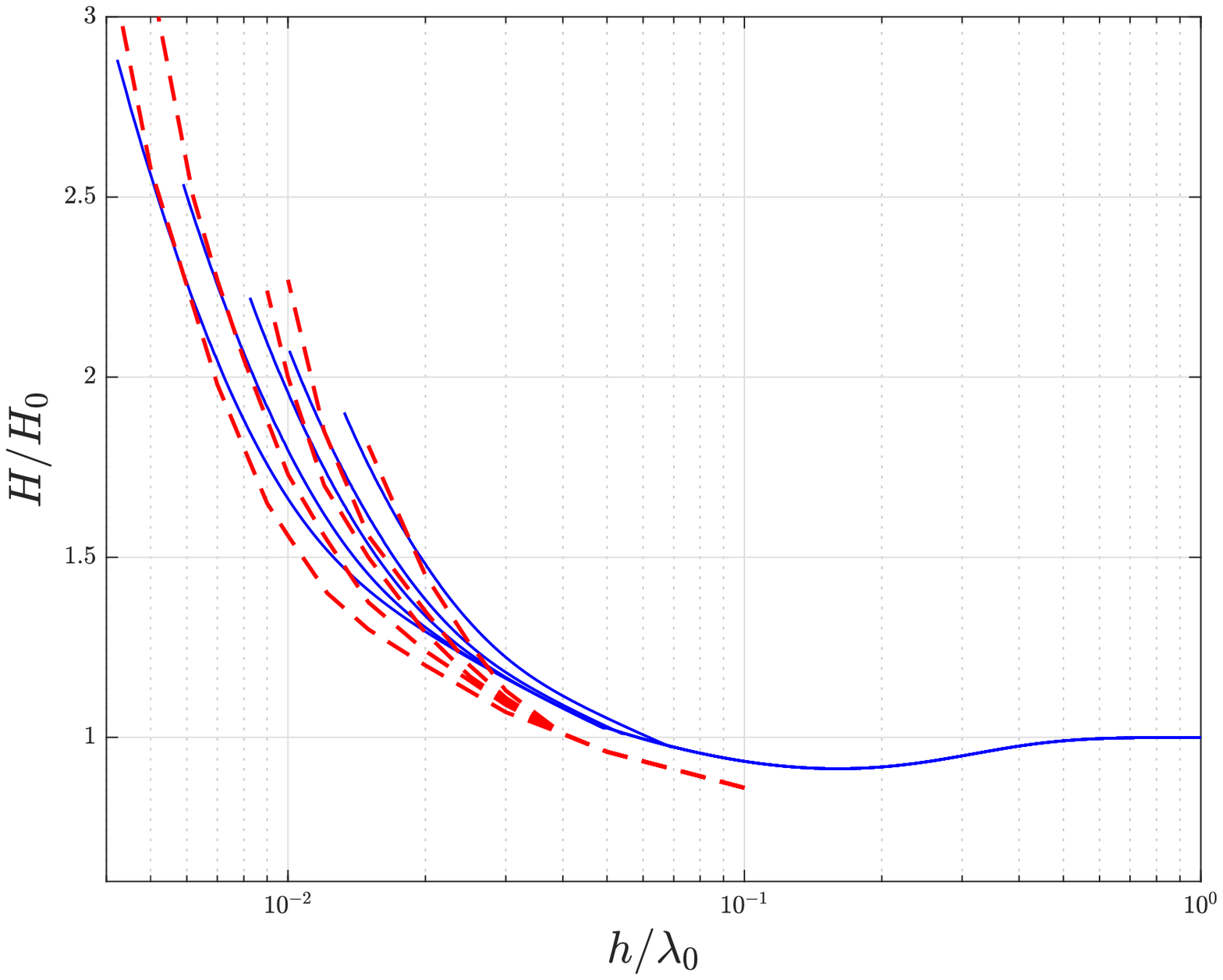}
  \label{SampleHu}
\end{minipage}
 \caption{\small Shoaling curves based on present theory in blue compared to the theory 
presented by Svendsen and Brink-Kj\ae r \cite{svendsen1972} in red. 
Deep water values $H_0/\lambda_0 = \{0.001, 0.002, 0.004, 0.006,0.01\}$.}
\label{ShoalingCurve}
\end{figure}

We note that the present implementation is able to determine the wave height further 
into the shoaling region as compared to the curves presented by \cite{khorsand2014shoaling}. 
This is due to the simplicity of the implementation, solving 
two nonlinear equations rather than three. 
We also see in Figure \ref{ShoalingCurve} that our theory is in fairly good agreement 
with the shoaling curves presented in \cite{svendsen1972}
which are in reasonable agreement with higher-order theories.
Moreover, it is clearly visible in Figure \ref{ShoalingCurve} that our curves
do not feature a discontinuity in wave height.

%
%

Finally, comparisons of shoaling curves computed using the method at hand
with data provided by wave tank experiments are presented. 
The experimental data are taken from \cite{svendsen1977wave} where considerable pains 
were taken to stay within the correct scaling regime.
The waves were generated with a piston-type wavemaker, then propagated over a flat bottom 
with still water depth of $36$ cm before shoaling on a $1:35$ beach.
%
%
%
%
\begin{figure}
  \includegraphics[scale=0.24]{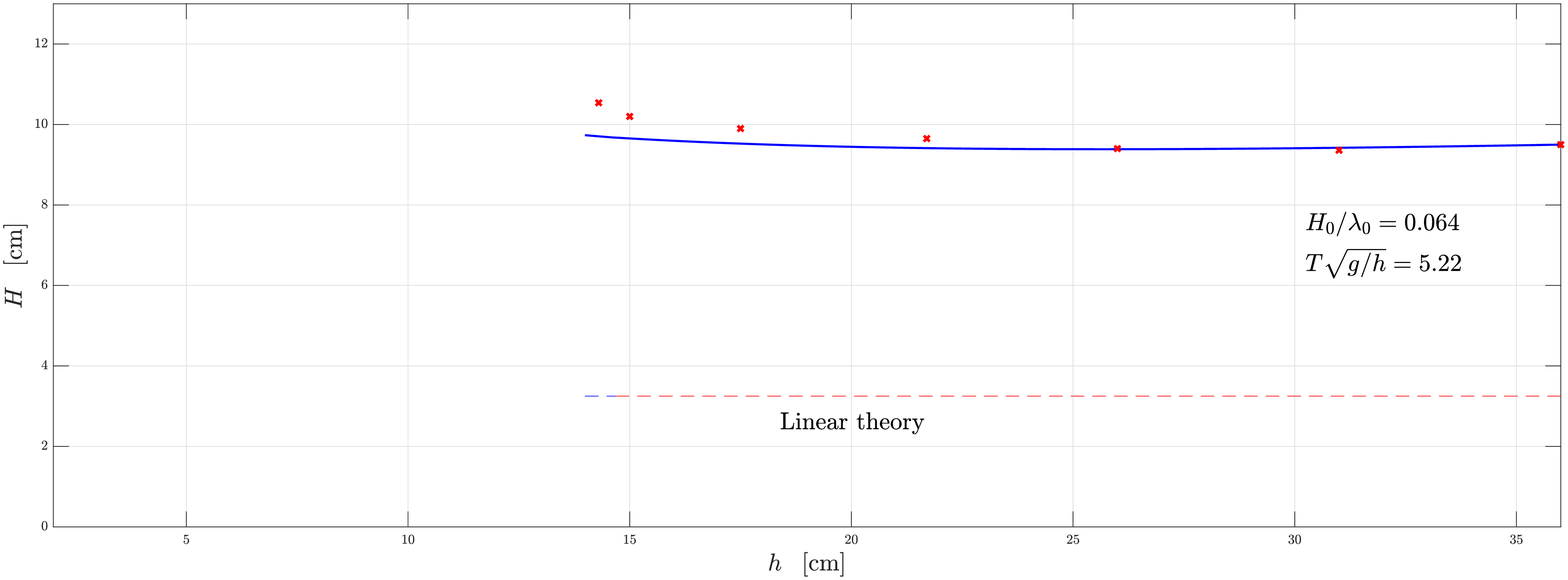}
  \centering
  \includegraphics[scale=0.24]{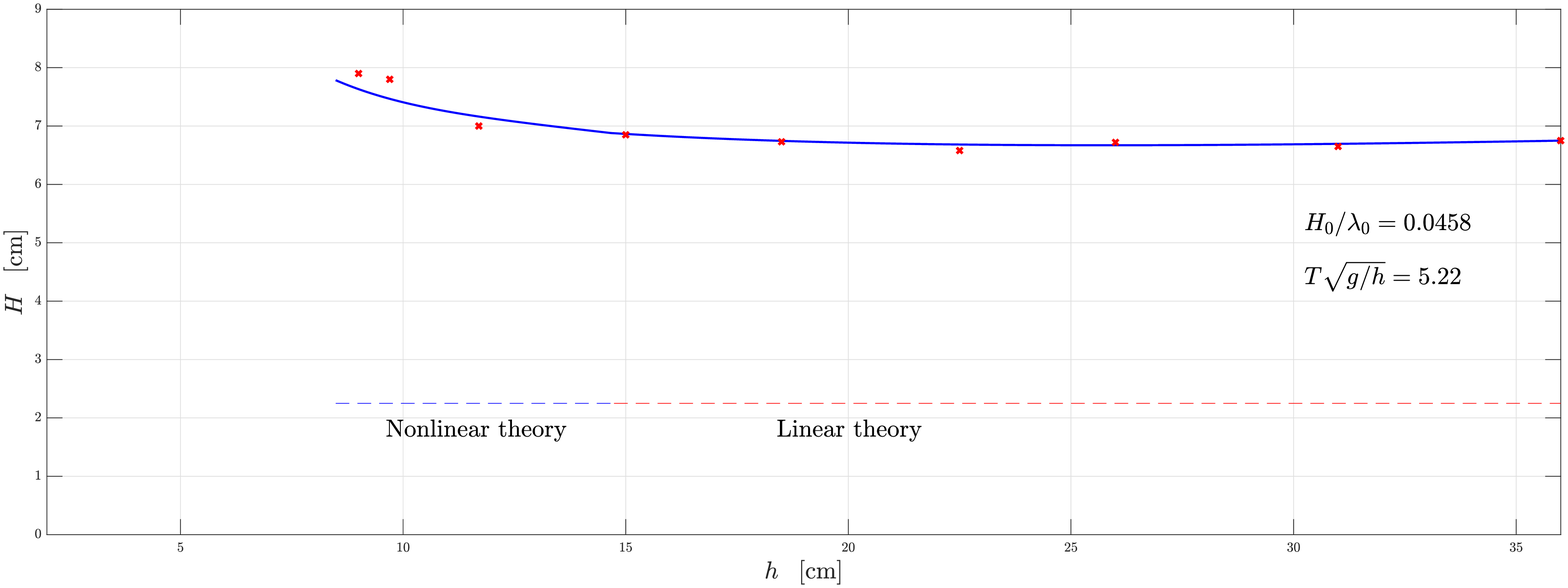}
  \centering
  \includegraphics[scale=0.24]{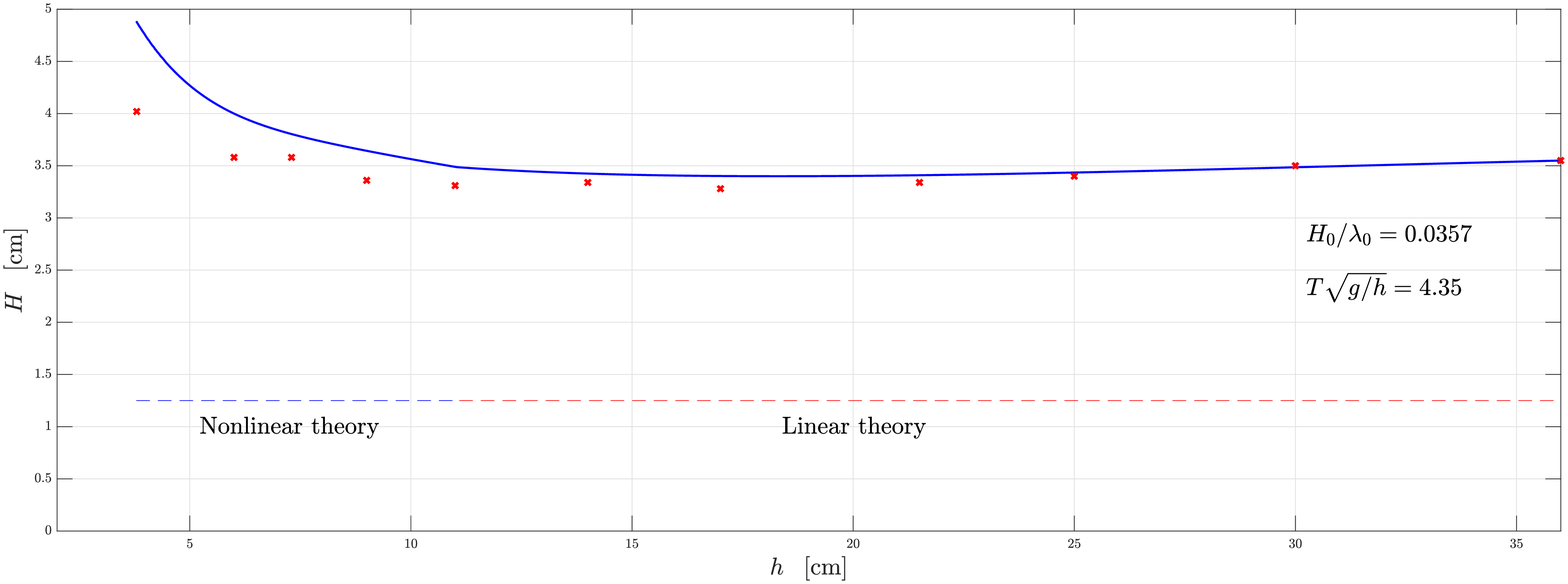}
  \centering
  \includegraphics[scale=0.24]{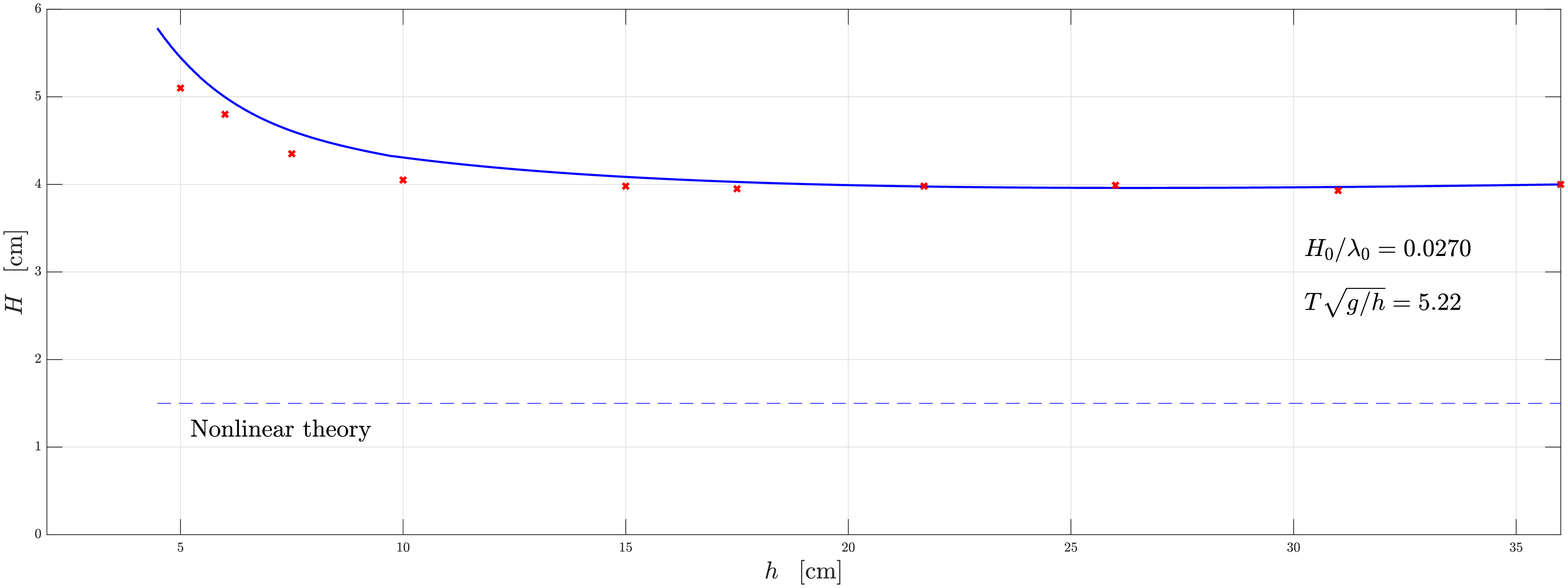}
  \centering
  \includegraphics[scale=0.24]{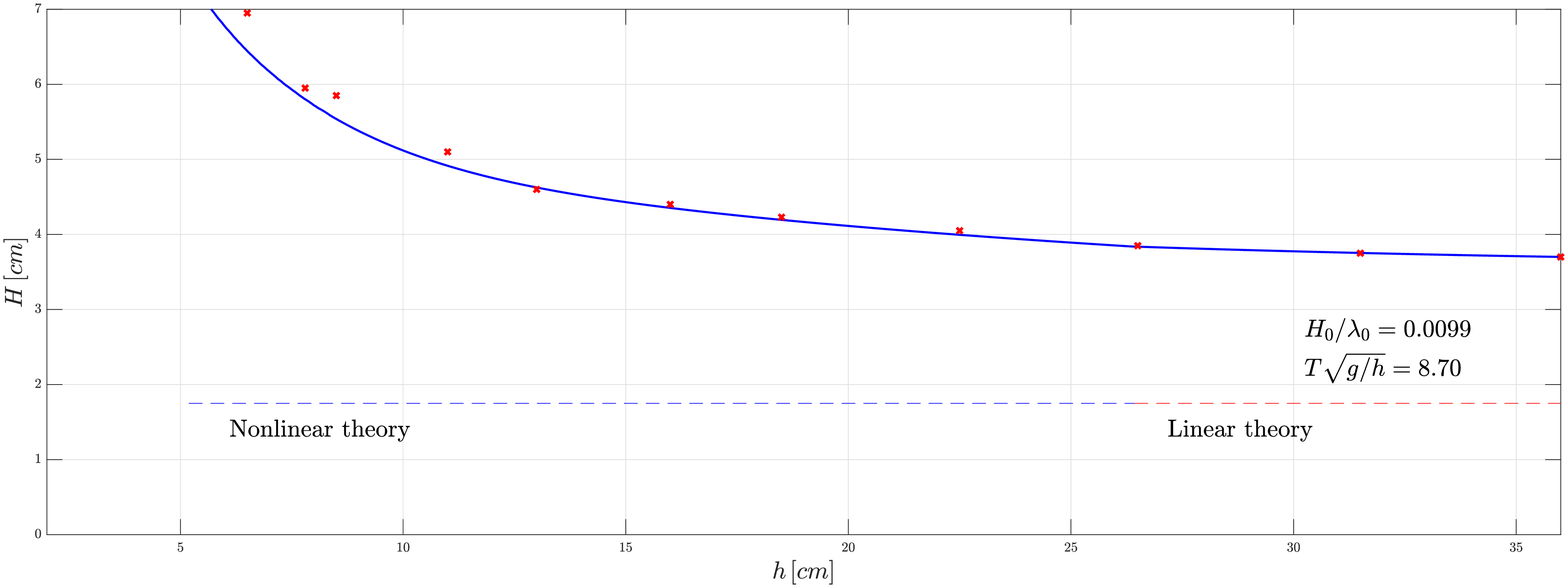}
  \centering
  \includegraphics[scale=0.24]{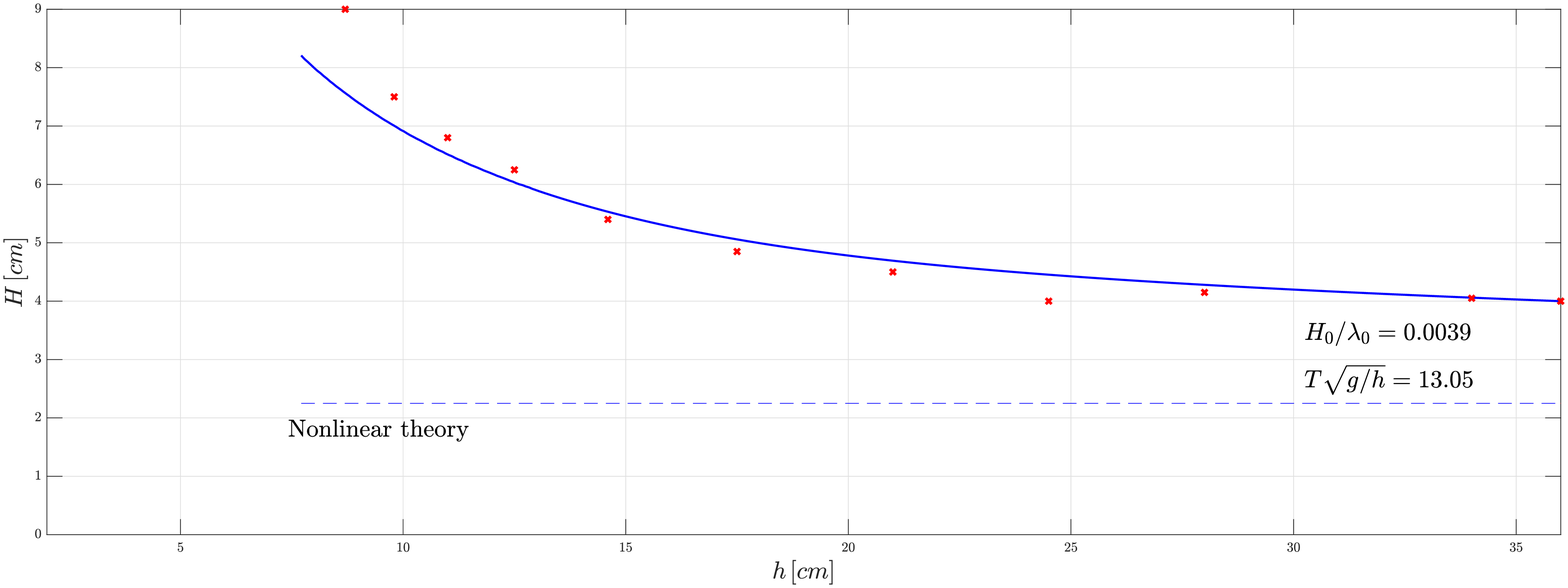}
  \centering
  \caption{\small Comparisons of shoaling curves from six experiments with a range 
                  of initial wave steepness.}
  \label{WaveFlumeShoaling}
\end{figure}
%
%
%
%
We initialize the code with wave height and wavelength at depth $h=36$cm at the toe of the slope.
The code then computes wave height $H$ with the linear model up to $h/\lambda_0>0.1$
at which time the code switches to the cnoidal theory.

We observe in Figure \ref{WaveFlumeShoaling} that there is very good agreement between 
the numerical model (shown in blue) and the experimental data (indicated by red crosses). 
The only exception is the experiment with deep water steepness $H_0/\lambda_0 = 0.064$,
shown in the uppermost panel in Figure \ref{WaveFlumeShoaling}.
This is a rather steep wave and it can be observed that even the linear
theory fails to get good agreement. This error is then propagated
and compounded by the nonlinear theory.
On the other hand, as long as the deep-water steepness is moderate enough,
the linear theory does an adequate job, and
we observe fairly good agreement in the plots between experimental data 
and numerical simulations for both linear and nonlinear regimes.
One should note that the higher-order theory by Cokelet was used numerically for the same 
data set in \cite{sakai1980wave}, but no better agreement was found than with the present theory.
We have also done some preliminary computations with a full Boussinesq model \cite{BOSZ},
and similar agreement with the wavetank data was found after some experimentation
with grid size and friction coefficients.

\section{Conclusion}
In this paper we have shown how to derive expressions for the energy flux and radiation stress
of a wavetrain in the context of the well known KdV equation. The derivations are based on
a general approach developed in \cite{AK2,AK4}, and yield approximations
of the same order as the KdV equation itself. Compared to previous definitions of such
quantities as for example in \cite{svendsen2006introduction,svendsen1972}, 
these expressions feature additional 
terms which are required by the theoretical underpinnings of the method used here.

The energy flux and radiation stress are used to develop shoaling equations for long waves
of small to moderate amplitude, and since such waves can be described by cnoidal functions,
the shoaling equations can be posed as a $3\times 3$ system of equations for the three
parameters of the cnoidal functions. The shoaling equations are then formulated in 
a numerically convenient way, and it is shown that they can be solved to yield
shoaling curves up to the breaking point. In particular, our formulation
resolves a problem encountered in earlier work \cite{khorsand2014shoaling}
where the curves terminated prematurely due to numerical instability.
The curves are compared with experimental data and higher-order theories,
and are found to be quite accurate. 
In addition, since the shoaling equations are
defined using approximations of energy flux and radiation stress
which are consistent with the approximations made in the KdV equation itself,
the shoaling curves are continuous, and no ad-hoc fixes such as advocated for in \cite{svendsen1977wave}
are necessary to get a continuous development of the wave height, 

Finally, since the radiation stress is included in the formulation of the shoaling equations,
it is possible to compute the set-down as part of the shoaling problem. Comparisons with
experimental data from \cite{bowen1968wave} show that the set-down can be computed
accurately. The nonlinear set-down defined here may also potentially be useful 
in the study of waves interacting with currents \cite{svendsen2006introduction}.

To put our study into context, note that due to the ready availability of
computational resources, shoaling is nowadays usually computed by utilizing
nearshore wave models which are used to forecast wave conditions in the coastal zone.
These models are generally known as Boussinesq models, and probably the first such model
was developed by Peregrine \cite{Peregrine1967}.
These models have become fairly sophisticated in recent years, being able
to treat even fairly short waves due to extending the dispersion properties
using ideas of Nwogu \cite{Nwogu} and Witting \cite{Witting}.
Some of the models currently in use are described in \cite{Brocchini2013,Madsen,BOSZ}.
In particular, as shown for example in \cite{MFW}, 
shoaling can be computed accurately with such models.
Following ideas of \cite{Serre}, some models have been extended to be able to 
treat larger-amplitude waves, and it is now possible 
to simulate waves with fully nonlinear and highly dispersive models \cite{LB,WeiKirby}
or with the full Euler equations \cite{Grilli94}.
Nevertheless, for practical purposes,
these models need to be initialized with data which are generally given
by stochastic wave models, and in general the transition between stochastic
and deterministic models is still poorly understood. 
The method of computing wave height and set-down put forward here
may have some potential if coupled with stochastic input data (see \cite{Holt})
since the computational complexity of our method is by far
smaller than for any phase-resolving nearshore model 
and no tuning is necessary with the current model.

\section*{Acknowledgments}
{\small
The authors thank Volker Roeber for helpful discussions.
This work has received funding from the European Union's {\em Horizon 2020} research and
innovation programme under grant agreement No. 763959.




}
\appendix
\section{Integrals of cnoidal functions}
In order to define the two shoaling models we need to 
handle integrals of various powers of derivatives of $\eta$. 
In particular we need to determine $\overline{\eta}$, $\overline{\eta^2}$, $\overline{\eta^3}$, 
$\overline{\eta_{xx}}$ and $\overline{\eta \eta_{xx}}$ in order to evaluate \eqref{setDown} 
in terms of $m$ and the current depth $h$. 
First note that we can write the time-averaged $\eta$ 
given by \eqref{CnoidalWaveSolution} in the more convenient form
\begin{equation*}
    \overline{\eta} = \frac{1}{T}\int_0^{T}\eta\Big{(}2K(m)( \frac{t}{T} - \frac{x}{\lambda})\Big{)} dt
    =
    f_2 + H
    \int_0^{1} \text{cn}^2(2K(m)\xi;m) d\xi,
\end{equation*}
\noindent
(for more details see \cite{svendsen2006introduction}). 
Similarly considerations apply to the powers of $\eta$
which involve integrals of the form
\begin{align*}
    & \int_0^{1} \text{cn}^2(2K\xi;m) d\xi = \frac{1}{4mK}\Big{(}E-(1-m)K\Big{)},\\
    \ 
    \\ 
    \ 
    & \int_0^{1} \text{cn}^4(2K\xi;m) d\xi = \frac{1}{3m^2}\Big{(}3m^2-5m+2(4m-2)\frac{E}{K}\Big{)},\\
    \ 
    \\ 
    \ 
    & \int_0^{1} \text{cn}^6(2K\xi;m) d\xi = \frac{1}{5m^2}\Big{(}4(2m^2 - 1)\int_0^{1} \text{cn}^4(2K\xi;m) d\xi + 3(1-m^2)\int_0^{1} \text{cn}^2(2K\xi;m) d\xi \Big{)}.
\end{align*}
%
%
These expressions can be found in \cite{lawden2013elliptic}. 
The terms $\overline{\eta_{xx}}$ and $\overline{\eta \eta_{xx}}$ also include such terms
due to the identity
\begin{equation*}
    \eta_{xx}(2K\xi;m) = 4K^2H(2 - 2m + (8m - 4)\text{cn}^2(2K\xi;m) - 6m \text{cn}^4(2K\xi;m)),
\end{equation*}
and similar relations which can be found in \cite{abramowitz1965handbook}. 
Combining the results above we deduce that

\begin{align*}
    & \overline{\eta}
    \hspace{5mm} = f_2 + H \int_0^{1} \text{cn}^2(2K\xi;m) d\xi,\\
    \ 
    \\ 
    \ 
    & \overline{\eta^2}
    \hspace{3.6mm} = f_2^2 + 2Hf_2\int_0^{1} \text{cn}^2(2K\xi;m) d\xi  + H^2\int_0^{1} \text{cn}^4(2K\xi;m) d\xi, \\
    \ 
    \\
    \ 
    & \overline{\eta^3}
    \hspace{3.6mm} = H^3\int_0^{1} \text{cn}^6(2K\xi;m) d\xi + 3H^2f_2\int_0^{1} \text{cn}^4(2K\xi;m) d\xi + 3Hf_2^2\int_0^{1} \text{cn}^2(2K\xi;m) d\xi + f_2^3, \\
    \ 
    \\ 
    \ 
    &\overline{\eta_{xx}}
    \hspace{2mm} = \frac{3H^2}{2m}(- 3m^2\int_0^{1} \text{cn}^4(2K\xi;m) d\xi + 4m^2\int_0^{1} \text{cn}^2(2K\xi;m) d\xi - 2\int_0^{1} \text{cn}^2(2K\xi;m) d\xi- m^2 + 1),\\
\end{align*}
\begin{align*}
    &\overline{\eta\eta_{xx}}
    \hspace{0mm} = 3H^2f_3\int_0^{1} \text{cn}^4(2K\xi;m) d\xi 
    -
    3H^2f_1\int_0^{1} \text{cn}^4(2K\xi;m) d\xi
    +
    \frac{3}{2}Hf_1f_2 
    -
    \frac{3}{2}Hf_2f_3 \\
    \ 
    \\ 
    \
    &\hspace{1cm}+
    \frac{3}{2}H^2f_1\int_0^{1} \text{cn}^2(2K\xi;m) d\xi 
    -
    \frac{3}{2}H^2f_3\int_0^{1} \text{cn}^2(2K\xi;m) d\xi 
    +
    6H^2f_1m^2\int_0^{1} \text{cn}^4(2K\xi;m) d\xi\\
    \ 
    \\ 
    \ 
    & \hspace{1cm} -
    6H^2f_3m^2\int_0^{1} \text{cn}^4(2K\xi;m) d\xi
    -
    \frac{3}{2}Hf_1f_2m^2 
    +
    \frac{3}{2}3Hf_2f_3m^2 
    -
    \frac{3}{2}H^2f_1m^2\int_0^{1} \text{cn}^2(2K\xi;m) d\xi\\
    \ 
    \\ 
    \
    & \hspace{1cm} +
    \frac{3}{2}H^2f_3m^2\int_0^{1} \text{cn}^2(2K\xi;m) d\xi
    -
    3Hf_1f_2\int_0^{1} \text{cn}^2(2K\xi;m) d\xi 
    +
    3Hf_2f_3\int_0^{1} \text{cn}^2(2K\xi;m) d\xi \\
    \ 
    \\ 
    \ 
    & \hspace{1cm}-
    \frac{9}{2}H^2f_1m^2\int_0^{1} \text{cn}^6(2K\xi;m) d\xi
    +
    \frac{9}{2}H^2f_3m^2\int_0^{1} \text{cn}^6(2K\xi;m) d\xi \\ 
    \ 
    \\ 
    \ 
    & \hspace{1cm}+
    6Hf_1f_2m^2\int_0^{1} \text{cn}^2(2K\xi;m) d\xi 
    -
    6Hf_2f_3m^2\int_0^{1} \text{cn}^2(2K\xi;m) d\xi \\
    \ 
    \\ 
    \ 
    & \hspace{1cm}-
    \frac{9}{2}Hf_1f_2m^2\int_0^{1} \text{cn}^4(2K\xi;m) d\xi
    +
    \frac{9}{2}Hf_2f_3m^2\int_0^{1} \text{cn}^4(2K\xi;m) d\xi. \\\\
\end{align*}

\end{document}